# Temporal Analysis of Functional Brain Connectivity for EEG-based Emotion Recognition

Ensieh Khazaei, Hoda Mohammadzade*

*Abstract*— **EEG signals in emotion recognition absorb special attention owing to their high temporal resolution and their information about what happens in the brain. Different regions of brain work together to process information and meanwhile the activity of brain changes during the time. Therefore, the investigation of the connection between different brain areas and their temporal patterns plays an important role in neuroscience. In this study, we investigate the emotion classification performance using functional connectivity features in different frequency bands and compare them with the classification performance using differential entropy feature, which has been previously used for this task. Moreover, we investigate the effect of using different time periods on the classification performance. Our results on publicly available SEED dataset show that as time goes on, emotions become more stable and the classification accuracy increases. Among different time periods, we achieve the highest classification accuracy using the time period of 140s-end. In this time period, the accuracy is improved by 4 to 6% compared to using the entire signal. The mean accuracy of about 88% is obtained using any of the Pearson correlation coefficient, coherence and phase locking value features and SVM. Therefore, functional connectivity features lead to better classification accuracy than DE features (with the mean accuracy of 84.89%) using the proposed framework. Finally, in a relatively fair comparison, we show that using the best time interval and SVM, we achieve better accuracy than using Recurrent Neural Networks which need large amount of data and have high computational cost.**

## I. INTRODUCTION

EMOTIONS play an important role in human decisions. Emotion recognition has numerous applications in the enhancement of human life and human ability [1]. This interdisciplinary field plays a critical role in the development of psychology, neuroscience, cognitive science, and computer science [2], [3]. In the field of computer science, automatic emotion recognition is used to improve human-computer interface [4], [5]. Other applications of emotion recognition include lie detection, behavior prediction and health monitoring. Among these various applications, human-computer interface is particularly important [3], [6]. In this application, machines get the ability to understand the emotional states of people and so can interact better with them.

There are different approaches in emotion recognition which can be divided into two categories. The first category is based on non-physiological signals such as facial expressions, body movements and intonation [3], [4], [5], [6], [7]. The second category is based on physiological signals such as electroencephalogram, electrocardiography, heart rate and respiration signals [3], [4]. Physiological signals provide more comprehensive and complex information and their results are more accurate [3], [4]. Among physiological signals, EEG signals receive more attention because they can better express brain states [6].

There are different models for expressing human emotions which can be generally divided into two categories: discrete basic emotions and continuous emotions. In the discrete model, emotions are classified into a set of discrete labels including six basic emotions: happiness, sadness, surprise, fear, anger, and disgust [8]. In the continuous model, emotions are expressed using two dimensions of valence and arousal [9]. Valence indicates how much an emotion is positive or negative, and arousal indicates how much a person is excited or indifferent [6], [10].

Up to now, a wide range of research has been done in the field of EEG-based emotion recognition. However, most recent research in the field of emotion recognition has focused on finding the strong connections, the important frequency bands and the important electrodes. Unfortunately, little research has been done to examine the significant time periods and the role of functional brain connectivity features in emotion classification which are the focus of this study. In this paper, we look for the optimal time period for emotion classification. Moreover, we examine the performance of recurrent neural networks to find temporal patterns in emotion classification.

Ensieh Khazaei and Hoda Mohammadzade are with Department of Electrical Engineering, Sharif University of Technology, Azadi Avenue, Tehran 11155-4363, Iran. (e-mail: ensieh.khazaei@ee.sharif.edu, hoda@sharif.edu).

* Corresponding author
.

The structure of this article is as follows. Section II gives a brief overview of related research on emotion recognition using EEG signals. Section III describes the dataset, preprocessing methods, feature extraction, and classification procedure. The results of recurrent neural networks and the optimal time period are given in Section IV. In Section V, we discuss our results and compare them against previous work. In Section VI, the conclusion is presented.

## II. RELATED WORK

In recent years, due to the development of dry electrode techniques [11], [12], [13], EEG signals, possessing high temporal resolution, are frequently used for emotion recognition. In this section, we review some of the previous research in the field of emotion recognition using EEG signals. In [2], differential entropy (DE), power spectral density (PSD), differential asymmetry (DASM), rational asymmetry (RASM) and differential causality (DCAU) in the five frequency bands, which are delta (1-3 Hz), theta (4-7 Hz), alpha (8-13 Hz), beta (14-30 Hz) and gamma (31-50 Hz), are extracted from the public SEED database [14] and then emotion classification is performed using deep neural networks. The results of this paper show that beta and gamma perform better than other frequency bands or in other words, there exist specific neural patterns in high frequency bands for positive, neutral, and negative emotions through time-frequency analysis. This paper also concludes that utilizing the electrodes that are located in the temporal area (FT7, FT8, T7, T8, C5, C6, TP7, TP8, CP5, CP6, P7, and P8) can increase classification accuracy by 2.66% compared to using all electrodes. In [3], by examining the DE and energy spectrum features extracted from their collected dataset, the authors infer that there is a close relationship between the emotional state of individuals and information of gamma band. In [5], a number of functional brain connectivity features like PLV, Pearson correlation coefficient and phase lag index (PLI) [15] are extracted and then the resulting connectivity matrices are fed to neural networks using different methods for spatial arrangement of electrodes. The purpose of this paper is to investigate the effect of spatial information in emotion classification. In [5], the arrangement of electrodes in the connectivity matrix is investigated and it is discovered to be effective in classification accuracy. In [16], the graph of PLV is obtained and then four features are extracted from the graph.

The authors then examine the simultaneous use of four graph features and local features such as DE and PSD.

They observed that adding graph features to local features leads to an increase in classification accuracy. Furthermore, this paper analyzes different frequency bands from which the PLV graph features are extracted, showing that utilizing the gamma and beta bands results in better performance compared to other frequency bands. In [17], the effect of familiar stimuli on emotion classification is investigated. The stimulus was a selected collection of music, and participants had education in music. Each participant listened to 8 familiar music and 8 unfamiliar music. Finally, results show that the classification accuracy is higher for unfamiliar music.

In [18], the stable patterns of EEG signal over time in emotion recognition are studied. The results show that

a) the lateral temporal areas are more active in positive emotions than negative ones in beta and gamma bands.

b) Brain activity in a neutral state has a greater alpha response in parietal and occipital regions.

c) Negative emotions are more active in delta band in the parietal and occipital regions.

d) Activity of gamma band in the prefrontal region is more in negative emotions.

Although a lot of works have been done on emotion recognition using EEG signals, a small number of them studied functional brain connectivity and the importance of different time periods during the stimuli which are the focus of this paper. The framework of our proposed approach is depicted in Fig. 1. First of all, we processed raw EEG signals and extracted some features from the preprocessed signal. Then a feature selection method was employed to decrease the number of features and select the most informative features. Finally, we did temporal analysis and recognized the emotion using the obtained information in temporal analysis step. All of these steps will be explained in details in Section III.

## III. METHODS

### A. Data Acquisition

We use the public SEED database [14] in this study. This database was recorded from 15 participants (7 males and 8 females, age range: 19-28 years). The stimuli in this database are 15 Chinese clips that create three emotions of positive, negative and neutral in participants; there are five clips for each emotion. The criteria for video selection were as follows: 1) the clips should not be too long to tire the participants, 2) the meaning of the videos can be understood without any explanation, 3) all the moments of the video convey only one emotion to the subject. The videos are approximately 4 minutes. The EEG signal of the participants was recorded using 62 electrodes placed on the head according to the 10-20 system [18]. EOG signal was recorded to remove the eye movement artifacts [2]. The location of the electrodes is shown in Fig. 2.

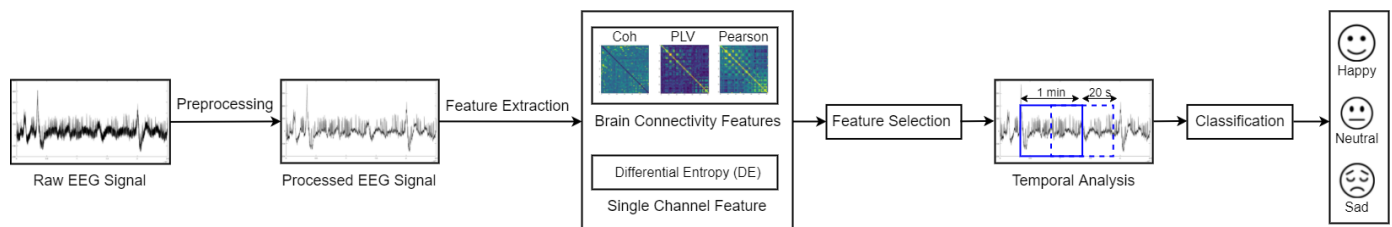

Fig. 1. The framework of our proposed approach

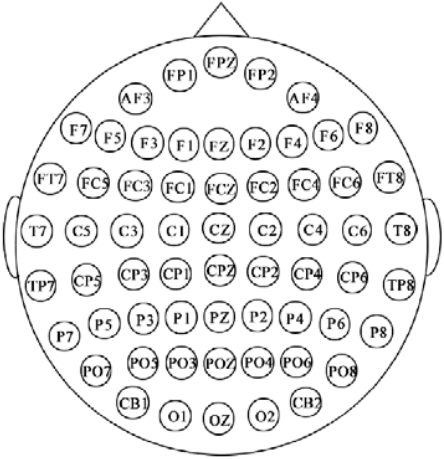

Fig. 2. Location of electrodes in SEED database [2]

The sampling rate of EEG recording was 1000 Hz. The signals were recorded in three sessions, where the time interval between two consecutive sessions was at least 7 days. The stimuli were the same in all three sessions; three sessions were recorded from each subject, and in each session, the participant watched all 15 video clips. Fig. shows detailed protocol in each session.

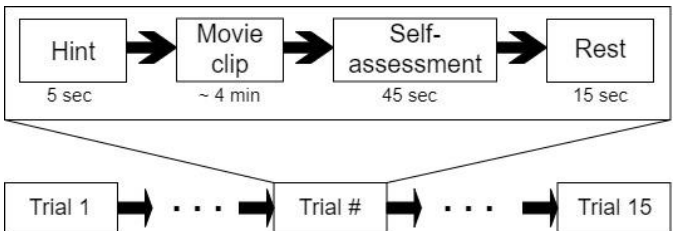

Fig. 3. Protocol of EEG recording in each session

The participants were alerted 5s before watching the video. After watching each video, an assessment form was given to the participant to express their emotions in 45s. 15s after the assessment was intended for rest [18].

### B. Preprocessing

The raw EEG signals are not appropriate for feature extraction and a series of preprocessing steps are needed beforehand. The raw EEG signals are down-sampled to 200 Hz. The EEG signals are visually checked in order to remove the parts of the signal that are heavily contaminated with EMG and EOG [2]. The EOG signals recorded during the experiment are also used to detect blink artifacts [19]. EEG signal are processed with a bandpass filter with frequency band between 0.3 Hz and 50 Hz [2]. Finally, the signals are divided into delta (1-4 Hz), theta (4-8 Hz), alpha (8-14 Hz), beta (14-31 Hz) and gamma (31-50 Hz) frequency bands and then in the time domain, the signal of each band is divided into time windows of 2s without overlapping. Time windowing is helpful to use the stationarity assumption in each time window because when the length of a time series is shorter, this assumption is more valid. Finally, features will be extracted from each of the 2s windows.

### C. Feature Extraction

In previous studies, various features such as DE, PSD, DASM, RASM, etc. have been used in emotion recognition. These features represent the local activity of electrodes and do not take into account the relationship between the electrodes. Among the local features, DE has the best performance in emotion classification [2], [6], [18]. Functional brain connectivity features show the relationship between different brain regions' activities which are recorded by different electrodes. Pearson correlation coefficient, coherence and PLV which are three important connectivity features are extracted in this study. These three features are non-directed and as a result their connectivity matrix is symmetric [20].

#### 1) Pearson Correlation Coefficient

Pearson correlation coefficient is a simple criterion of linear correlation between two time series. This criterion calculates the linear relationship between the two signals $x$ and $y$ as follows:

$$\rho_{xy} = \frac{E[(x-\mu_x)(y-\mu_y)]}{\sigma_x \sigma_y} \\ = \frac{1}{N}\sum_{n=1}^{N}\frac{(x(n)-\mu_x)(y(n)-\mu_y)}{\sigma_x \sigma_y} \quad (1)$$

Where $N$ is the number of time samples, $\mu_x$, $\mu_y$ are mean of the signals $x$ and $y$, respectively and $\sigma_x$, $\sigma_y$ are standard deviation of the signals $x$ and $y$, respectively. This criterion has a value between -1 and +1. The value of +1 in this criterion means positive correlation of two signals and the value of -1 means negative correlation. The value of zero means that the two signals are not correlated.

#### 2) Coherence

Coherence indicates a linear correlation between two signals in the frequency domain. First, the cross-spectral density function $S_{xy}(f)$, is calculated as

$$S_{xy}(f) = \sum_{\tau=-\infty}^{+\infty} E[x_n y_{n+\tau}] e^{-j2\pi f \tau} \quad (2)$$

Where $x_n$ and $y_{n+\tau}$ are the $n$-th sample of signal $x$ and the $(n+\tau)$-th sample of signal $y$, respectively. Coherence is then calculated using the cross-spectral density function as follows:

$$C_{xy}(f) = \frac{S_{xy}(f)}{\sqrt{S_{xx}(f)S_{yy}(f)}} \quad (3)$$

$$COH_{xy}(f) = |C_{xy}(f)| \quad (4)$$

The cross-spectral density function of the signals $x$ and $y$ is normalized using the spectral density function of signal $x$ and signal $y$. The value of coherence is between zero and +1. The value of +1 means the highest correlation at that frequency. Coherence with the value of zero means independence at that frequency.

*3) Phase Locking Value*

The most common feature for measuring phase synchronization between different areas of the brain is Phase Locking Value (PLV). Suppose two signals $x_1$ and $x_2$ are filtered with a bandpass filter. An analytical signal $z_i$ can be defined as

$$z_i(t) = x_i(t) + jH(x_i(t)) = A_i(t)e^{j\phi(t)} \quad (5)$$

where $H$ is the Hilbert transform operator. The phase difference is calculated using

$$\Delta\phi(t) = \arg(\frac{z_1(t)z_2^*(t)}{|z_1(t)||z_2(t)|}) \quad (6)$$

PLV is then defined as [16]:

$$\text{PLV} = \left| \frac{1}{N} \sum_{j=0}^{N-1} e^{j\Delta\phi(t)} \right| \quad (7)$$

PLV is between zero and one. The value of one indicates a phase lock, and the value of zero indicates a random phase distribution over time. The amount of PLV is independent of the signal amplitude and depends only on the phase difference between two signals [22].

*D. Classification*

Most of the works that have used the SEED database have considered the first 9 trials of each session as training data and the final 6 trials of each session as test data. Due to the small number of trials in this database, it is better to use Leave-One-Out cross validation strategy. Leave-One-Out cross validation strategy decreases the probability of overfitting and has a higher degree of validity. In this strategy, we consider one trial as test and the other 14 trials as training. We repeat this procedure until every trial is considered once as test. Finally, we average the classification accuracies over all folds. We use SVM with linear kernel for classification and also the emotion classification is subject dependent. The number of samples is equal to *number of time windows * number of trials * number of sessions*. The number of time windows depends on the length of time windows. In this study, the length of time window is 2s, and time windows are non-overlapping. Given that our final purpose is to label the trials while samples are time windows, we use a simple voting scheme to determine the label of a trial using the label of its time windows ; for instance, if there are $n$ time windows in a trial, we will have a vector of length $n$ of labels and then, we vote among all the labels in the vector to find the label of the trial.

*E. Dimension Reduction*

It is known that an imbalance between the length of the feature vector and the number of samples can cause overfitting. The length of feature vector for local features such as DE is equal to the number of channels which is equal to 62 in SEED database. Coherence, PLV, and Pearson correlation coefficient are non-directed, and their connectivity matrix is symmetric. Therefore, the upper or lower half of connectivity matrix is sufficient for classification. As a result, the length of the vector of connectivity features is equal to (62*61)/2=1891, which is large relative to the number of samples which is approximately 4700. Thus, we should use dimension reduction techniques to prevent overfitting. In this study, we use Fisher score to reduce the dimension. The Fisher score assigns a score to each feature where higher score for a feature shows that the feature is more discriminative for classification. The score of a feature is obtained as follows [23]:

$$F(i) = \frac{\sum_{k=1}^{c}(\bar{x}_{k,i} - \bar{x}_i)^2}{\sum_{k=1}^{c} \frac{1}{n_k - 1} \sum_{j=1}^{n_k}(x_{k,j,i} - \bar{x}_{i,k})^2} \quad (8)$$

where $\bar{x}_i$ is the mean of the *i*-th feature in all samples, $x_{k,j,i}$ is the *i*-th feature of the *j*-th sample in the *k*-th class, $\bar{x}_{i,k}$ is the mean of the *i*-th feature in the *k*-th class and $n_k$ represents the number of samples in the *k*-th class.

*F. Recurrent Neural Networks*

A recurrent neural network (RNN) is a type of neural networks in which connections between nodes form a graph along a sequence [25]. RNNs [26] are widely used in text analysis [27], [28], text generation [29], [30], speech recognition [31], time series prediction [32] and processing of time signals such as electroencephalogram [33]. The main idea of RNNs is the use of hidden states which are used as the network memory and are responsible to store the information from past inputs. In RNNs, the output $y_t$ at each time $t$ is calculated by the hidden state $h_t$ at time $t$. The hidden state $h_t$ at time $t$ is also updated with the input $x_t$ at time $t$ and the hidden state $h_{t-1}$ at time $t-1$. Mathematically, there are two important equations at each time step in RNNs as follows [25]:

$$h_t = \sigma(W_{hx}x_t + W_{hh}h_{t-1} + b_h) \quad (9)$$

$$y_t = \text{softmax}(W_{yh}h_t + b_y) \qquad (8)$$

$W_{hx}$, $W_{hh}$ and $W_{yh}$ are the weight matrices. $b_h$ and $b_y$ are bias parameters which are utilized to learn the offset and $\sigma$ is the sigmoid function.

These networks need a large number of samples for training. The number of our samples in SEED database is small. Therefore, we select Gated Recurrent Units (GRUs) [34] which are a cell type of RNNs and their few parameters make them a suitable choice for small datasets [35]. In addition, we increase the number of samples by adding Gaussian noise with standard deviations of 0.001, 0.004, 0.008 and 0.012 to the samples according to [4]. The utilized network consists of a GRU layer that has 16 units and a fully connected layer for emotion classification. The filtered signal is then divided into time sliding windows of 180s with step length of 2 seconds as the input of network. We choose 180s as the length of our inputs because the signals in SEED database have different length but their minimum length of signals is 180s. We also opt step length of 2s to cover the end part of signals and also increase the number of inputs of the network. Leave-One-Out cross validation strategy is chosen for calculating accuracy in this section.

## IV. RESULTS

### A. Investigation of frequency bands

Previous research in the field of emotion recognition has shown that higher frequency bands have more information about emotional states and gamma band has the best performance in the emotion classification [2], [3], [6], [16], [18], [24]. However, most of them do not investigate the performance of different frequency bands for functional brain connectivity features. In this research, in addition to DE features, we investigate the performance of different frequency bands for coherence, PLV and Pearson correlation coefficient. Table I shows the classification accuracy for our features in different frequency bands before dimension reduction.

TABLE I
THE MEAN ACCURACIES AND STANDARD DEVIATIONS (%) OF DIFFERENT FREQUENCY BANDS

|        | Delta       | Theta       | Alpha       | Beta        | Gamma       |
|--------|-------------|-------------|-------------|-------------|-------------|
| DE     | 64.44/18.97 | 58.67/15.97 | 65.33/18.72 | **81.32/16.17** | 80.44/16.03 |
| Coh    | 40.44/17.36 | 51.11/21.03 | 66.67/19.52 | 77.78/18.97 | **79.56/17.54** |
| PLV    | 26.67/17.81 | 40.89/21.21 | 60.44/22.17 | 75.56/18.11 | **76.45/17.97** |
| Pearson| 35.11/10.83 | 49.78/19.00 | 65.78/19.97 | 75.56/17.02 | **78.22/18.25** |

As seen in Table I, the gamma and beta bands have better classification accuracy rather than other bands. Specifically, DE achieves its maximum accuracy in beta band while functional brain connectivity features perform best in gamma band. The classification accuracy of DE feature in gamma band is 0.88% less than beta band while in functional connectivity features, the accuracy of gamma band is at least 0.89% more than beta band. Moreover, our focus in this research is on functional brain connectivity features, so we select gamma band for further analysis. We also do dimension reduction for all features only on the gamma band.

### B. Dimension Reduction Results

We apply the Fisher score for DE, Pearson correlation coefficient, coherence and PLV features in gamma band because in Section IV, we investigate the performance of different frequency bands and we observe that the gamma band has better performance in emotion recognition. We calculate F-score for all features and then sort them in order to find the first 100, 200, 300, 400, 500, 600, 700, 800, 1200 selected feature for coherence, PLV and Pearson correlation coefficient. We also calculate F-score for DE features and then sort them in order to find the first 10, 20, 30, 40, 50 selected features for DE. The optimal number of selected features for DE, Pearson correlation coefficient, coherence and PLV are 40, 700, 200 and 400, respectively. Classification accuracy versus the number of functional connectivity features and DE are shown in Fig. 4 and Fig. 5, respectively. Fig. 4 and Fig. 5 show the importance of feature selection; If the number of selected features is little, then the features will not be discriminative and the classification accuracy will be low. On the other hand, if the number of features is too large, the accuracy of classification is reduced because overfitting happens. Table II shows the classification accuracy before and after performing Fisher on our features.

TABLE II
THE MEAN ACCURACIES AND STANDARD DEVIATIONS (%) BEFORE AND AFTER DIMENSION REDUCTION

|               | DE          | Coh         | PLV         | Pearson     |
|---------------|-------------|-------------|-------------|-------------|
| Before fisher | 80.44/16.03 | 79.56/17.54 | 76.45/17.97 | 78.22/18.25 |
| After fisher  | 80.90/16.87 | 84.00/15.49 | 82.22/16.65 | 82.67/14.21 |

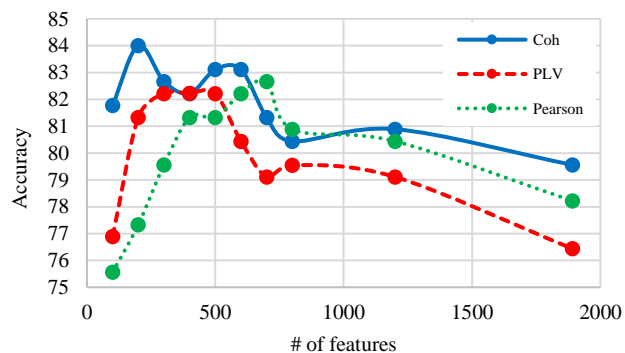

Fig. 4. Average accuracy of classification over different subjects versus the number of selected functional connectivity features

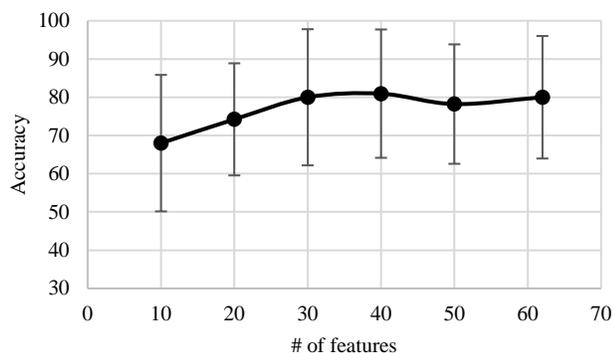

Fig. 5. Average accuracy of classification over different subjects versus the number of selected DE features with Fisher

### C. Emotion Classification Using Fusion of DE and Functional Brain Connectivity

According to our results in the previous part and a wide range of research in the field of emotion recognition, we carry out our experiments only in gamma band. It is worth mentioning that from this part to the end of this study, we utilize all features in gamma band after dimension reduction for our analysis. In this part, we investigate the effect of the fusion of functional connectivity and DE features on the classification accuracy in emotion recognition. Table III shows classification accuracy using decision-level fusion of functional connectivity and DE features.

TABLE III
THE MEAN ACCURACIES AND STANDARD DEVIATIONS (%) OF FUNCTIONAL CONNECTIVITY FEATURES AND FUSION OF FUNCTIONAL CONNECTIVITY FEATURES WITH DE FEATURE

|  | Coh | PLV | Pearson |
| --- | --- | --- | --- |
| Brain connectivity | 84.00 / 15.49 | 82.22 / 16.65 | 82.67 / 14.21 |
| Brian connectivity + DE | 84.00 / 15.28 | 83.11 / 13.58 | 84.44 / 13.72 |

The classification accuracy using the DE feature in gamma band is equal to 80.90% with 16.87% std. Table III shows that the fusion of functional connectivity and DE features improves the classification accuracy. However, because of our focus on temporal pattern of functional connectivity features, we do not fuse the functional connectivity and DE features in the rest of this article to purely observe the temporal pattern of functional brain connectivity features. It is also worth mentioning that we investigated the fusion of different connectivity features together but no improvement was observed.

### D. Analysis of Temporal Pattern

Recent works mostly investigated location and frequency of brain activity during emotional stimuli. To the best of the authors' knowledge, brain activity during an emotional stimulation interval has not been yet studied in the literature. The length of stimulus in SEED database is approximately 4 minutes. In this section, we want to find temporal pattern over different time periods during each trial. To this end, we consider one minute sliding window with step length of 20 seconds and calculate classification accuracy in each sliding window. Table IV shows the classification accuracy in different time periods of signals.

As seen in Table IV, there is a temporal pattern in the classification accuracy in which, as time goes on, the mean accuracy increases. The best time period is 140s-end in which the mean accuracy is maximum and standard deviation is minimum. Comparing the results of Table III and Table IV shows that the mean accuracy in 140s-end interval is increased compared to the entire signal by 3.99%, 4%, 6.22% and 5.77% for DE, coherence, PLV and Pearson correlation coefficient, respectively. Standard deviation in 140s-end interval is also decreased compared to the entire signal.

It is noteworthy that the length of stimuli in many datasets for EEG-based emotion recognition such as DEAP [36] are less than two minutes while the temporal analysis of SEED shows the importance of brain activities that occur after two minutes from the onset of stimuli. Therefore, we conclude that the appropriate length of stimuli in emotion recognition is about 3 to 4 minutes so that not only the subjects do not get tired but also there are the best periods for emotion recognition. In this study, we watched the videos and confirmed that all part of the videos express the same emotion as the label of the video. Although we can understand the label of the video from the beginning part of it, over time the emotion becomes more stable in our mind, and consequently the final intervals have better performance in emotion classification. We can also observe this phenomenon in our ordinary life, for instance, when a person tells a funny sentence, you feel happy and if that person tells a sequence of funny sentences, you feel happier and even a few minutes after telling those funny sentences, you still feel happy. Therefore, it seems that the length of stimuli in emotion recognition has an optimal interval at which the subject reaches the peak of his emotions. If the length of stimuli is less than the optimal value then, the subject's emotions are not well expressed, and if the length of stimuli is greater than the optimal length, the subjects get tired.

TABLE IV
THE MEAN ACCURACIES AND STANDARD DEVIATIONS (%) OF DIFFERENT TIME INTERVALS

|  | 0-60s | 20-80s | 40-100s | 60-120s | 80-140s | 100-160s | 120-180s | 140s-end |
| --- | --- | --- | --- | --- | --- | --- | --- | --- |
| DE | 65.33 / 20.65 | 64.89 / 18.07 | 69.78 / 17.43 | 72.89 / 16.61 | 77.33 / 16.09 | 81.33 / 14.29 | 81.78 / 11.39 | **84.89 / 13.20** |
| Coh | 61.33 / 22.70 | 69.33 / 18.82 | 74.22 / 19.00 | 77.78 / 15.46 | 82.67 / 13.75 | 86.67 / 15.11 | 86.22 / 15.82 | **88.00 / 13.38** |
| PLV | 59.56 / 22.17 | 68.44 / 19.75 | 72.00 / 19.38 | 76.00 / 16.67 | 81.33 / 16.36 | 85.33 / 16.56 | 85.33 / 13.84 | **88.44 / 10.53** |
| Pearson | 60.44 / 20.84 | 69.78 / 19.00 | 72.89 / 18.59 | 77.33 / 15.69 | 81.78 / 166.22 | 86.22 / 15.82 | 87.11 / 13.20 | **88.44 / 12.20** |

The results shown in Table IV are the averaged classification accuracies over different subjects. Another approach is to obtain the best interval for each subject separately. By using the second approach, we found that the best interval is not the same for all subjects, however, the best interval is 140s-end for 13, 11, 10 and 11 subjects (out of 15 subjects) in DE, coherence, PLV and Pearson correlation coefficient features, respectively. The best interval for other subjects occurs 120s, 100s, 80s and 100s after the onset of stimulation for DE, coherence, PLV and Pearson correlation coefficient features, respectively. To visualize the results of the temporal analysis, the accuracy using each time interval is shown in Fig. 6, separately for each feature. In this figure, squares with higher intensity are indicators of better classification accuracy.

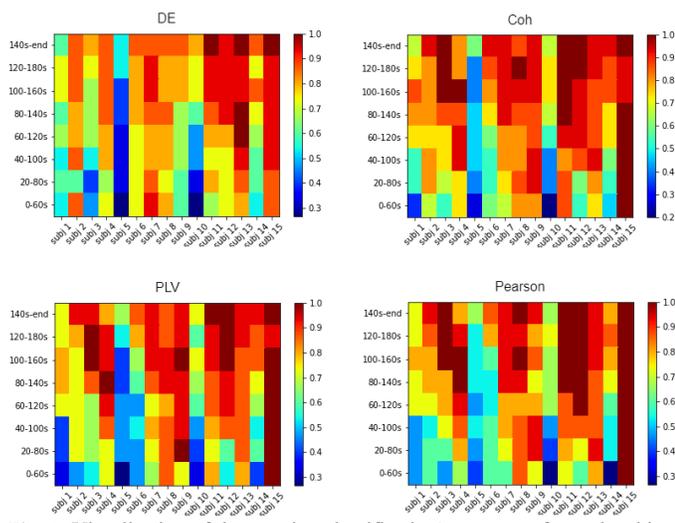

Fig. 6. Visualization of the emotion classification's accuracy for each subject in different time intervals for DE, coherence, PLV, and Pearson correlation coefficient

Table V shows the comparison between results of these two approaches: 1) averaging between subjects and then finding the best interval (first row of Table V), 2) finding the best interval for each subject and then averaging between the maximum accuracy of each subject (second row of Table V).

TABLE V
THE MEAN ACCURACIES AND STANDARD DEVIATIONS (%) OF APPROACH 1) AVERAGING BETWEEN SUBJECTS AND THEN FINDING THE BEST INTERVAL, APPROACH 2) OBTAIN THE OPTIMAL INTERVAL FOR EACH SUBJECT

|            | DE            | Coh           | PLV           | Pearson       |
|------------|---------------|---------------|---------------|---------------|
| Approach 1 | 84.89 / 13.20 | 88.00 / 13.38 | 88.44 / 10.53 | 88.44 / 12.20 |
| Approach 2 | 86.22 / 11.94 | 92.00 / 11.32 | 91.56 / 10.53 | 91.11 / 10.59 |

The results of Table V show that the average accuracy in the second approach compared to that in the first approach increases by 1.33% to 4%. Therefore, regarding to different character and reaction of different subjects, it is a wise idea to find the best time interval for each subject independently and then predict the subject's emotion.

### E. Classification Using Recurrent Neural Network

In the previous section, we found the best time interval using SVM. In this section, we investigate the performance of recurrent neural networks [25] in finding the temporal pattern of brain activity during emotional stimulation. In order to provide a fair comparison, we augment the training set by adding Gaussian noise to the training samples and then adding them to the training set. We also employ GRUs with few parameters for compatibility with the size of the dataset. Data augmentation and reduction of the number of the parameters of the network prevent overfitting of the classifier to train samples. Table VI compares the results of the recurrent neural network with the results of the previous section obtained using SVM.

TABLE VI
THE MEAN ACCURACIES AND STANDARD DEVIATIONS (%) OF SVM AND GRU FOR FUNCTIONAL CONNECTIVITY FEATURES

|     | Coh           | PLV           | Pearson       |
|-----|---------------|---------------|---------------|
| SVM | **88.00 / 13.38** | **88.44 / 10.53** | **88.44 / 12.20** |
| GRU | 84.00 / 11.37 | 87.11 / 12.46 | 81.78 / 11.94 |

The results of Table VI show that the performance of SVM is better than recurrent neural networks for the utilized dataset in this study. In addition, neural networks require large amounts of data and their computational cost is high. Therefore, the proposed time-interval selection method provides both advantages of relatively high accuracy and low computational cost for the task of emotion recognition.

## V. DISCUSSION

In this study, our aim was to examine the functional connectivity features in more detail as well as the temporal analysis of the brain's response to emotional stimuli, which have been rarely studied in the field of emotion recognition. In this section, we compare our results with previous works. However, it is difficult to compare the classification accuracy of our method with previous works because they have used different datasets with different number of subjects and various methods of representing emotions. Moreover, even in the works that use SEED dataset for their experiments, the first 9 trials are considered as training and the final 6 trials as test data. As a result, due to the small size of this dataset, the accuracy reported by those methods may not be much reliable. In order to obtain more reliable results, we used the Leave-One-Out strategy on the trials to calculate the classification accuracy for each subject. Consequently, our classification accuracy might be lower than some of those reported in the literature due to our different strategy for constructing train and test sets but it is more valid. Therefore, in this comparison, it is important to note that firstly, we used a more reliable method to measure the classification accuracy, and secondly, our goal was to find optimal time intervals and analyze temporal patterns, but not to propose a method to increase the classification accuracy in emotion recognition.

TABLE VII
THE ACCURACY OF EMOTION CLASSIFICATION IN PREVIOUS WORKS

| Study | Year | Data | Features | Frequency band | Classifier | Number of classes | Description | Accuracy |
|---|---|---|---|---|---|---|---|---|
| Zheng et al [2] | 2015 | SEED | DE | Gamma band | DBN | 3 | | 79.19% |
| Duan et al [3] | 2013 | Collected | DE | Gamma band | SVM | 2: positive and negative emotions | | 84.25% |
| Wang et al [4] | 2018 | SEED | DE | Concatenation of all frequency bands | ResNet | 3 | - | 75% |
| Zheng et al [18] | 2019 | SEED | DE | Concatenation of all frequency bands | GELM | 3 | - | 91.07% |
| Song et al [6] | 2020 | SEED | DE | Gamma band | GCNN | 3 | - | 83.36% |
| Our method | - | SEED | DE | Gamma band | SVM | 3 | - | 84.89 % |
| Moon et al [5] | 2018 | DEAP | PLV, PCC | Concatenation of all frequency bands | CNN (two-layer) | 2: low and high valence | | 96.62%, 93.80% |
| Li et al [16] | 2019 | SEED | ENP (extracted from PLV graph) | Gamma band | SVM | 3 | - | 33%* |
| Wu et al [38] | 2019 | SEED | Three topological features (for PCC and Coh indices) | Concatenation of all frequency bands | SVM | 3 | Feature-level fusion method was used to fuse topological features | 79.16%, 78.15% |
| Our method | - | SEED | Coh, PLV, Pearson | Gamma band | SVM | 3 | - | 88.00%, 88.44%, 88.44% |

PCC stands for Pearson correlation coefficients.
* They achieved 79% classification accuracy using GELM.

We summarize the accuracy rates of emotion classification in previous works in Table VII. In [2], [3], [6], and [18], frequency analysis for some features like DE and PSD have been investigated, but the functional connectivity features have not been investigated except in [16]. In [16], it has been shown that using the gamma and beta bands results in better performance compared to using the other frequency bands for emotion classification. As discussed in Section IV.A, the results of our frequency analysis on the functional connectivity features using SEED database confirm the results of previous articles. It can be concluded that these two frequency bands contain more discriminative information related to the emotional states of individuals' brain.

Regarding temporal analysis, unfortunately, none of the previous works has studied the temporal patterns during stimuli, but we examined this for the first time. We observed that each subject has an optimal time interval in which the classification performance is higher than the accuracy of using the whole signal. It is noteworthy that, as discussed in Section IV.D, the optimal time interval for all subjects in this database occurs 80 seconds after the onset of stimuli. Therefore, the different subjects behave relatively similar to each other.

Using DE features extracted from the gamma band we obtained the mean accuracy of 84.89%. More details regarding the comparison between our result and the previous works are as follow. In [2], the accuracy of 79.19% using DE features extracted from the gamma band and Deep Belief Networks (DBN) as the classifier has been achieved. In [18], the mean accuracy of 91.07% using the concatenation of DE from all frequency bands and Graph regularized Learning Machine (GELM) has been obtained. Despite the good accuracy obtained in [18], the computational complexity of GELM is higher than SVM which is utilized in our study. In [3], the accuracy of 84.25% using DE features from the gamma band has been achieved. In [3], the authors collected their own dataset which contains only two positive and negative emotional states, while in our research, the number of classes is three. In [6], the average accuracy of 83.36% using DE features from the gamma band and Graph Convolutional Neural Networks (GCNNs) has been obtained on SEED dataset. Despite of the good performance of Graph Neural Networks, their computational cost is higher than SVM which is used in our work. In [4], the maximum accuracy of 75% has been reached using DE features and ResNet [37] on SEED dataset. Due to the small size of SEED dataset, in [4] first data augmentation methods are used to provide sufficient data for training of the network.

Using functional connectivity features, we have achieved the mean accuracy of 88%, 88.44%, and 88.44% for coherence, PLV, and Pearson correlation coefficients, respectively. More details regarding the comparison between our results and the previous works are as follow. In [5], the accuracy rates of 93.80% and 96.62% using a two-layer Convolutional Neural Network (CNN) as the classifier and Pearson correlation coefficients and PLV matrices as features have been achieved, respectively. In [5], only the results of high and low valence classification are reported, which are on DEAP dataset. They have not reported their accuracy No results on the arousal classification have been reported. In [16], 33% accuracy on SEED dataset using SVM classifier and ENP features extracted from the gamma band using PLV graph has been achieved. In [38], first of all, the Pearson correlation coefficients and coherence were extracted then three topological features of strength, clustering coefficient, and eigenvector centrality were calculated for each connectivity indices. By using the feature-level fusion method, these three topological features were concatenated. Finally, they obtained the accuracy of 79.16% and 78.15% on Pearson correlation coefficients and coherence indices.

It can be seen that although there are a few works with better accuracy rates on SEED dataset, the accuracy that we have obtained using the optimal time intervals and a simple SVM classifier is competitive with the results of previous works while it has less computational cost.

## VI. CONCLUSION

In this study, we examined the applicability of functional brain connectivity for the emotion recognition task. Our results indicate that using functional connectivity features better emotion classification accuracy can be obtained compared to using DE features. In order to study the temporal variation of different features in terms of classification accuracy, we used a one-minute sliding window on the signals. Classification accuracy increased with sliding window progress leading us to the conclusion that using the 140s-end interval results in the best performance compared to using other intervals as well as compared to using the entire signal in SEED dataset. The mean accuracy and standard deviation using this interval (140s-end) for DE, PLV, coherence and Pearson correlation coefficient are 84.89%/13.20%, 88.44%/10.53%, 88%/13.38% and 88.44%/12.20%, respectively, showing 4-6% improvement compared to using the entire signal. Although the temporal behavior of each subject is different, the selected interval (140s-end) is the best period for at least two thirds of the subjects. The results show that by using the best time interval, we can achieve high accuracy by relatively low computational cost and limited number of training samples.

## DECLARATION

The authors declare no conflict of interest.

## FUNDING

This research did not receive any specific grant from funding agencies in the public, commercial, or not-for-profit sectors.